# Chaotic Spin-Wave Solitons in Magnetic Film Feedback Rings

Zihui Wang,[1] Aaron Hagerstrom,[1] Justin Q. Anderson,[2] Wei Tong,[1,3] Mingzhong Wu,[1]* Lincoln D. Carr,[2] Richard Eykholt,[1] and Boris A. Kalinikos[1,4]

[1]*Department of Physics, Colorado State University, Fort Collins, Colorado 80523, USA*

[2]*Department of Physics, Colorado School of Mines, Golden, Colorado 80401, USA*

[3]*High Magnetic Field Laboratory, Chinese Academy of Sciences, Hefei, Anhui 230031, China*

[4]*St.Petersburg Electrotechnical University, 197376, St. Petersburg, Russia*

Chaotic spin-wave solitons in magnetic film active feedback rings were observed for the first time. At some ring gain level, one observes the self-generation of a single spin-wave soliton pulse in the ring. When the pulse circulates in the ring, its amplitude varies chaotically with time. Numerical simulations based on a gain-loss nonlinear Schrödinger equation reproduce the observed responses.

If one amplifies the output signal from a dissipative transmission line and then feeds it back to the input of the line, one creates an active feedback ring system – a driven damped system. In the steady state of this system, the energy loss of the wave in the dissipative line is compensated by the energy gain provided by the amplifier. Examples of this type of ring system include fiber ring lasers [1,2,3], magnetic film feedback rings [4,5,6,7], and electromagnetic transmission line oscillators [8,9]. These systems are excellent test beds for studies of nonlinear dynamics and, therefore, have attracted considerable works. The two main focus areas are (1) *envelope solitons* and (2) *chaos*. For (1), the main work has been on the demonstration of envelope solitons in a wide variety of ring configurations [2,3,5,7,8]. For (2), the main focus has been on the use of various configurations to demonstrate chaotic excitations through different nonlinear processes [1,2,3,4,6,8,9] and the development of new models to describe chaotic excitations in certain ring systems [1-3].

Recent theoretical work discovered that the active feedback system could also support *chaotic solitons.* Specifically, Soto-Crespo *et al*. [10], Zhao *et al*. [11], and Karar *et al*. [12] reported optical envelope solitons that circulated in fiber feedback rings and had their amplitudes varying with time in a chaotic manner. This discovery opened a completely new paradigm in the field of nonlinear science. This is because solitons and chaos are usually considered to exist in opposite physical regimes and present two totally unconnected aspects of nonlinear dynamics [13,14,15]. In spite of the significance of this discovery, however, the experimental demonstration of such chaotic solitons has been rather limited. The only demonstration so far was done by Zhao *et al*. for optical soliton pulses in fiber feedback rings [11]; however, neither the solitonic nature of the pulses nor the chaotic nature of their behavior was confirmed.

This Letter reports on the first experimental demonstration and modeling of chaotic spin-wave solitons in magnetic film active feedback rings. As the ring gain is increased to a certain threshold level, one observes the self-generation of a single spin-wave envelope soliton that circulates in the ring with constant amplitude. With a further increase in the ring gain, this soliton pulse develops into a chaotic soliton - a soliton whose amplitude changes chaotically with time. The pulse has a hyperbolic secant shape and a flat phase profile across its width, which are the signatures of a soliton. The overall time-domain signal resulting from the circulation of the pulse shows a finite correlation dimension and a positive Lyapunov exponent, which are clear evidence of chaos. Numerical simulations, based on a gain-loss nonlinear Schrödinger equation (GLNLS), reproduced the observed responses. At relatively low ring gain levels, there is even a quantitative agreement between the numerical and experimental results.

It is important to emphasize that, although these results were obtained for a magnetic film feedback ring, the work has implications for other driven damped nonlinear systems, including optical fiber rings [1-3] and electromagnetic transmission line oscillators [8,9]. It is also important to highlight that this work demonstrates a new type of chaotic microwave pulse generator. Chaotic microwave sources are critically needed by chaotic radar [16] and chaotic communications [17].

The feedback ring consisted of a magnetic yttrium iron garnet (YIG) film strip and two microstrip transducers placed over the YIG strip to excite and detect spin waves [7]. The output signal from the detection transducer was fed back to the excitation transducer through a microwave amplifier and an adjustable microwave attenuator. The YIG strip was magnetized by a static magnetic field which is parallel to the YIG strip length. This film-field configuration supports the propagation of backward volume spin waves along the YIG strip and, at the same time, prohibits the three-wave interactions of such waves [18,19]. The ring signal was sampled through a directional coupler, with feeds to a spectrum analyzer for frequency analysis and an oscilloscope for temporal signal measurements. For the data presented below, the YIG strip was 5.6 μm thick, 2.1 mm wide, and 52 mm long. The magnetic field was 938 Oe. The microstrip transducers were 50 μm wide and 2 mm long elements. The transducer separation was held at 5.5 mm. The microwave amplifier had a peak output power of 2 W and a linear response over 1-8 GHz.

The feedback ring can have a number of resonance eigenmodes that exhibit low decay rates [7]. The frequencies of these modes can be determined by the phase condition $k(\omega)l+\phi_e=2\pi n$, where $k$ is the spin-wave wavenumber, $\omega$ is the frequency, $l$ is the transducer separation, $\phi_e$ is the phase shift introduced by the electronic circuits, and $n$ is an integer. At a low ring gain $G$, all eigenmodes experience an overall net loss, and there is no spontaneous signal in the ring. If the ring gain is increased to a certain level, here taken as $G=0$, the eigenmode with the lowest decay rate will start to self-generate in the ring and one will obtain a continuous wave response. A further increase in $G$ leads to the excitation of additional modes and a comb-like frequency spectrum, which, in the time domain, corresponds to a spin-wave soliton that circulates in the ring; and then to the broadening of each mode in the frequency spectrum, which, in the time domain, corresponds to the realization of chaotic solitons as reported below. Here, both the excitation of new modes and the mode broadening are realized through four-wave interactions. At even higher gain levels, one obtains the circulation of two or more spin-wave pulses in the ring.

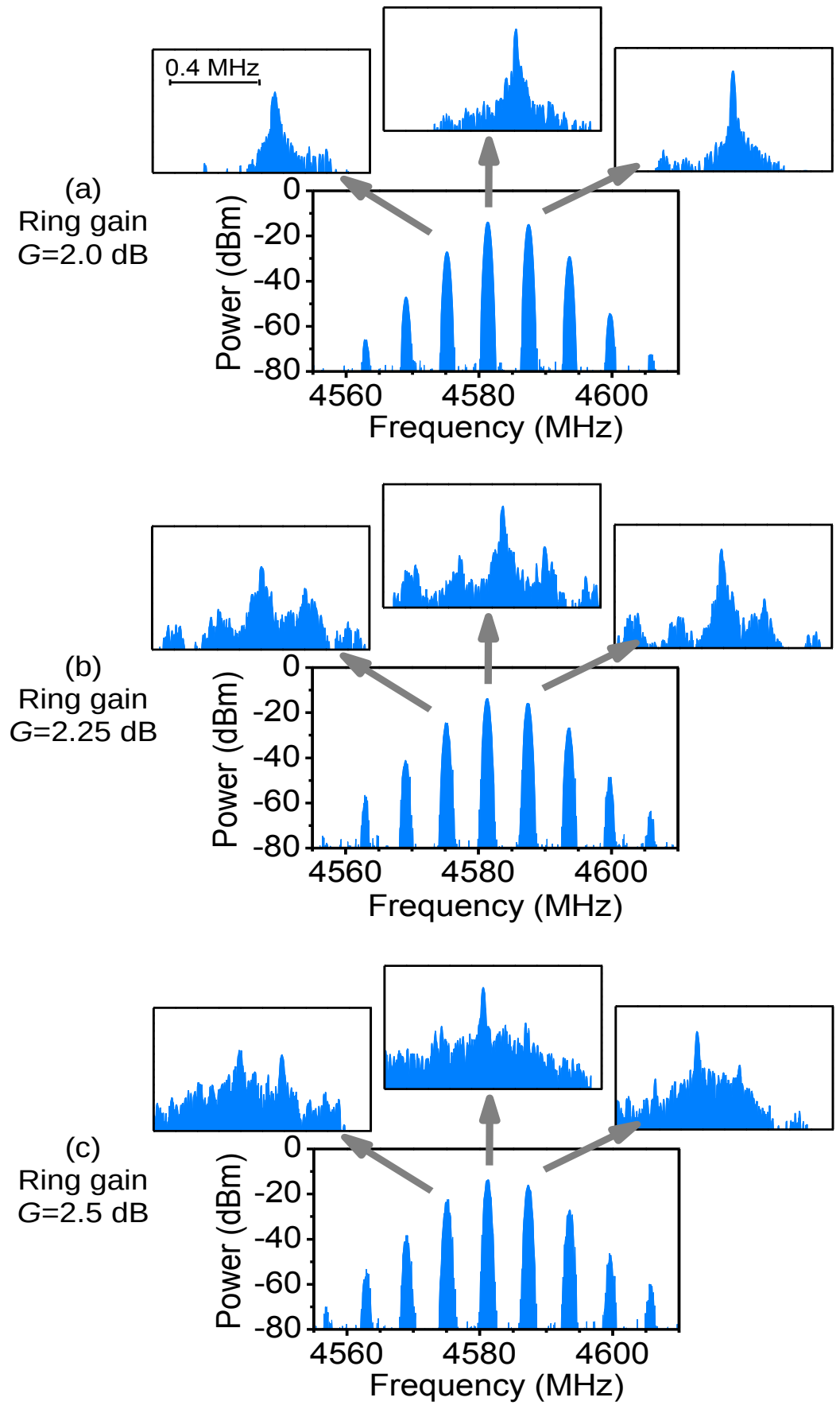

FIG. 1. Spectra for ring signals obtained at different ring gain ($G$).

Figure 1 shows representative power spectra for ring signals obtained at different ring gains. In each panel, the bottom diagram shows the full spectrum, and the top diagrams present ×32 expanded views for the three main peaks in the bottom diagram. All diagrams have the same vertical power scale. The three top diagrams for the same main peak also have the same frequency scale. The peak widths in all the diagrams are instrument limited.

The data in Fig. 1 demonstrate three results. (1) On a large frequency scale, as shown in all the bottom diagrams, the power spectrum has a comb-like structure. With an increase in $G$, this comb spectrum remains the same, except that there is a weak growth in mode intensity. (2) On a smaller frequency scale, each mode consists of a narrow single peak for $G$=2.0 dB. With an increase in $G$, one observes the excitation of new sideband peaks near the initial single peak, as shown in the top diagrams in (b), and then the wash out of those modes and the realization of broad spectra, as shown in the top diagrams in (c). (3) There is a slight shift of the modes to lower frequencies. This shift agrees with the fact that backward volume spin waves have a negative nonlinearity coefficient [18].

The time-domain signal obtained at $G$=2.0 dB consists of a uniform train of pulses. This signal corresponds to the clean comb spectrum in Fig. 1(a) and results from the continuous circulation of a single spin-wave soliton in the ring [7]. With an increase in *G* to 2.25 dB, the train becomes chaotic – the amplitude of the pulse varies chaotically with time. This corresponds to the excitation of new side modes shown in Fig. 1(b) and is the onset of chaos. With a further increase in *G*, one observes stronger chaotic behavior.

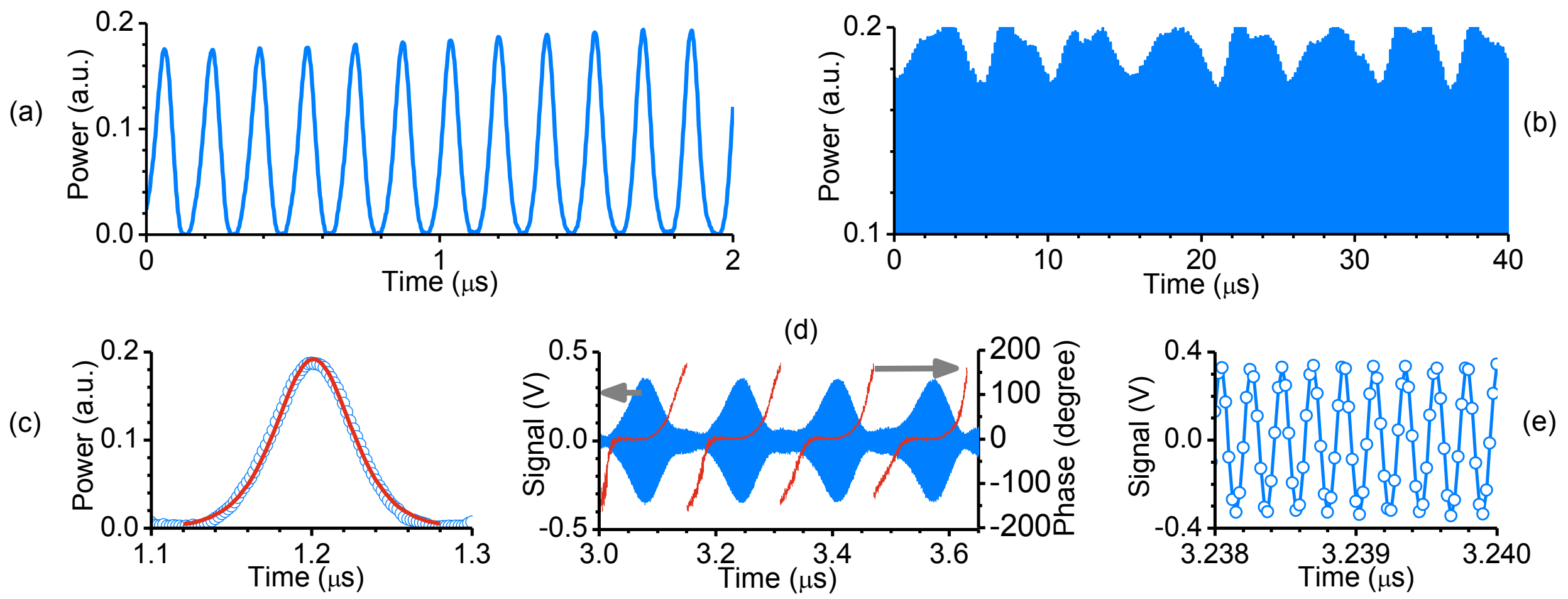

FIG. 2. Time-domain signal obtained at *G*=2.5 dB: (a)-(c) power profiles; (d)-(e) carrier waves. In (c), the circles show the actual data and the curve shows a hyperbolic secant function fit. The curves in (d) show the relative phase profiles of the corresponding pulses.

Figure 2 shows the time-domain signal obtained at *G*=2.5 dB. Graphs (a) and (b) show the power profile of the signal in different time and power scales. Graph (c) shows one pulse in (a) in an expanded time scale. The circles are data, and the curve is a fit to a hyperbolic secant squared function. Graph (d) shows the carrier waves (left axis) and phase profiles (right axis) for four pulses. Each phase profile shows the phase of the carrier wave relative to a reference continuous wave whose frequency was given by the main frequency of the carrier wave of the pulse. Graph (e) shows the carrier wave of one pulse in (d) in an expanded time scale.

The data in Fig. 2(a)-(b) show a train of chaotic pulses. This train corresponds to the circulation of a single spin-wave soliton pulse whose amplitude changes chaotically with time. The data in (c) shows a perfect hyperbolic secant function fit. The data in (d) show flat phase profiles across the central portions of the pulses. These results clearly confirm the solitonic nature of the pulses. The waveform in (e) and the clean phase profiles in (d) show that, in spite of the chaotic variation in amplitude, the soliton has a coherent carrier wave as a conventional soliton.

Figure 3 shows representative data that confirm the chaotic nature of the time-domain signals. Graph (a) shows a 3D attractor. Graph (b) shows plots of correlation sum *C* vs. probing distance *r* for embedding dimensions *m*=2-20. Graphs (c) and (d) show the correlation dimension and maximal Lyapunov exponent, respectively, as a function of *m*. The squares in (c) are for the *G*=2.25 dB signal. All other data in Fig. 3 are for the *G*=2.5 dB signal. The approaches for attractor construction and correlation sum calculation are the same as in Ref. [20]. The calculation of the maximal Lyapunov exponents involved the following steps [21]: (1) construction of the attractor; (2) identification of the nearest neighbor point to each of the points on the attractor; (3) examination on how these points separate as time increases; (4) average of the logs of the separations for a given time; (5) plotting of the average as a function of time; and (6) determination of the slope of the linear region in the plot. The obtained slopes were taken to be the maximal Lyapunov exponent $\lambda$.

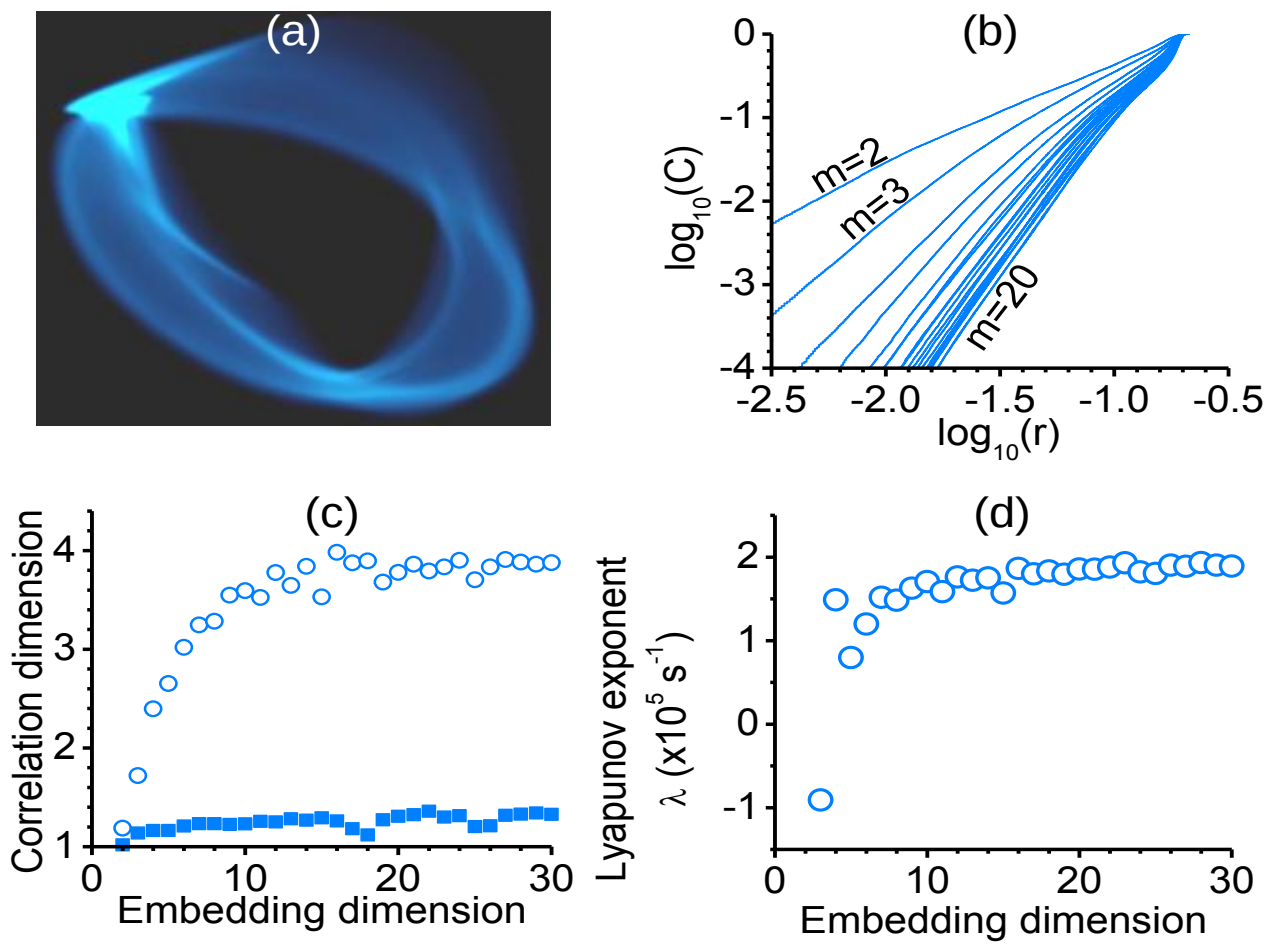

FIG. 3. Chaotic characterization of time-domain ring signals.

The attractor in Fig. 3(a) is smooth and has a visible structure. The correlation plots in (b) all show a linear regime, in which the slopes of the plots yield the dimension data shown in (c). The data in (c) clearly demonstrate saturation behavior and indicate a fractal dimension of about 1.27 for the $G$=2.25 dB signal and a higher dimension of about 3.83 for the $G$=2.5 dB signal. The response in (d) shows a saturation of $\lambda$ at about $1.9\times10^5$ $s^{-1}$. These results clearly confirm the chaotic nature of the measured signals. Note that the anomalously negative value of $\lambda$ for $m$=3 is due to the fact that the embedding dimension $m$ is less than the fractal dimension of the attractor.

Numerical modeling was performed with the following GLNLS equation:

$$i\frac{\partial u}{\partial t}=\left[-\frac{D}{2}\frac{\partial^2}{\partial x^2}+iL+(N+iC)|u|^2+(S+iQ)|u|^4\right]u, \quad (1)$$

where $u$ is a unitless spin-wave amplitude, $D$ is the dispersion, $N$ and $S$ are the cubic and quintic nonlinearity, respectively, $t$ is the 'temporal' evolution coordinate, $x$ is the 'spatial' coordinate of propagation boosted to the group velocity of the envelope, and $L$, $C$, and $Q$ are the linear, cubic, and quintic gains (if positive) or losses (if negative), respectively. Note that similar equations have been used to model exciton-polariton Bose-Einstein condensates [22] and mode-locked lasers [23].

The measurements indicate that nonlinearity and dispersion are the dominant sources of envelope shaping for spin waves and that the losses present in the ring are fully compensated by the amplifier. This imposed two constraints on modeling. (1) The coefficients $N$ and $D$ must be orders of magnitude larger than $L$, $C$, and $Q$. (2) The linear amplifier must compensate both the linear and nonlinear losses present in the film, requiring a net averaged linear gain ($L$>0). The dissipative terms represent the net gain and loss processes occurring in the ring averaged over several round trip times. One expects the use of this approximation to be valid when the time scale of envelope modulation is greater than the soliton round trip time. This condition is met by all ring gains in this work.

Simulations were performed using adaptive time-step Runga-Kutta for 'temporal' evolution and pseudospectral techniques for 'spatial' propagation. Periodic boundary conditions mimicked the propagation of a single soliton around the ring. Experimentally measured $D$ and $N$ values were used to generate a stable soliton train which then numerically propagated in the YIG strip with higher order nonlinearity. The cubic dissipation represented nonlinear loss processes ($C$<0). $Q$>0 was used to saturate the nonlinear loss. All simulations were run with $\max(|u|^2) < 1$ and the quintic nonlinearity was evaluated separately as a higher order nonlinearity ($S$ < 0) and as a high-power saturation of cubic nonlinearity ($S$ > 0). Finite correlation dimensions were observed numerically only for $S$ > 0. The amplitude of envelope modulation was seen to vary with the magnitude of $S$, while increases in gain quickly destroyed the solitonic nature of the pulse.

Figure 4 illustrates typical simulation data. The left and right columns show data for chaotic solitons with amplitude variations of 2.0% and 5.1%, respectively. These two variation levels are chosen because they match the experimental variations of the $G$=2.25 dB and $G$=2.5 dB signals, respectively. The simulation used the following parameters: $N$ = -9.0×10$^9$ rad/s, $D$ = 4.5×10$^3$ cm$^2$rad/s, $L$ = 5.9×10$^5$ rad/s, $C$ = -5.9×10$^5$ rad/s, and $Q$ = 5.9×10$^5$ rad/s. The value of $S$ was taken as 6.0×10$^9$ rad/s for the data with 2.0% variation and 12.0×10$^9$ rad/s for the data with 5.1% variation. The top and middle rows give power profiles in different time scales, which show trains of chaotic solitons just like those in Fig. 2 (a)-(c). The two graphs in the bottom row show the correlation dimension data. The left one indicates a fractal dimension of about 1.26, which closely matches that of the $G$=2.25 dB signal. The right one indicates a dimension of about 1.66, which is lower than that of the $G$=2.5 dB signal. Overall, one can see that the simulations reproduced the measured responses in terms of the amplitude variation, qualitative structure, and correlation dimension at low gains.

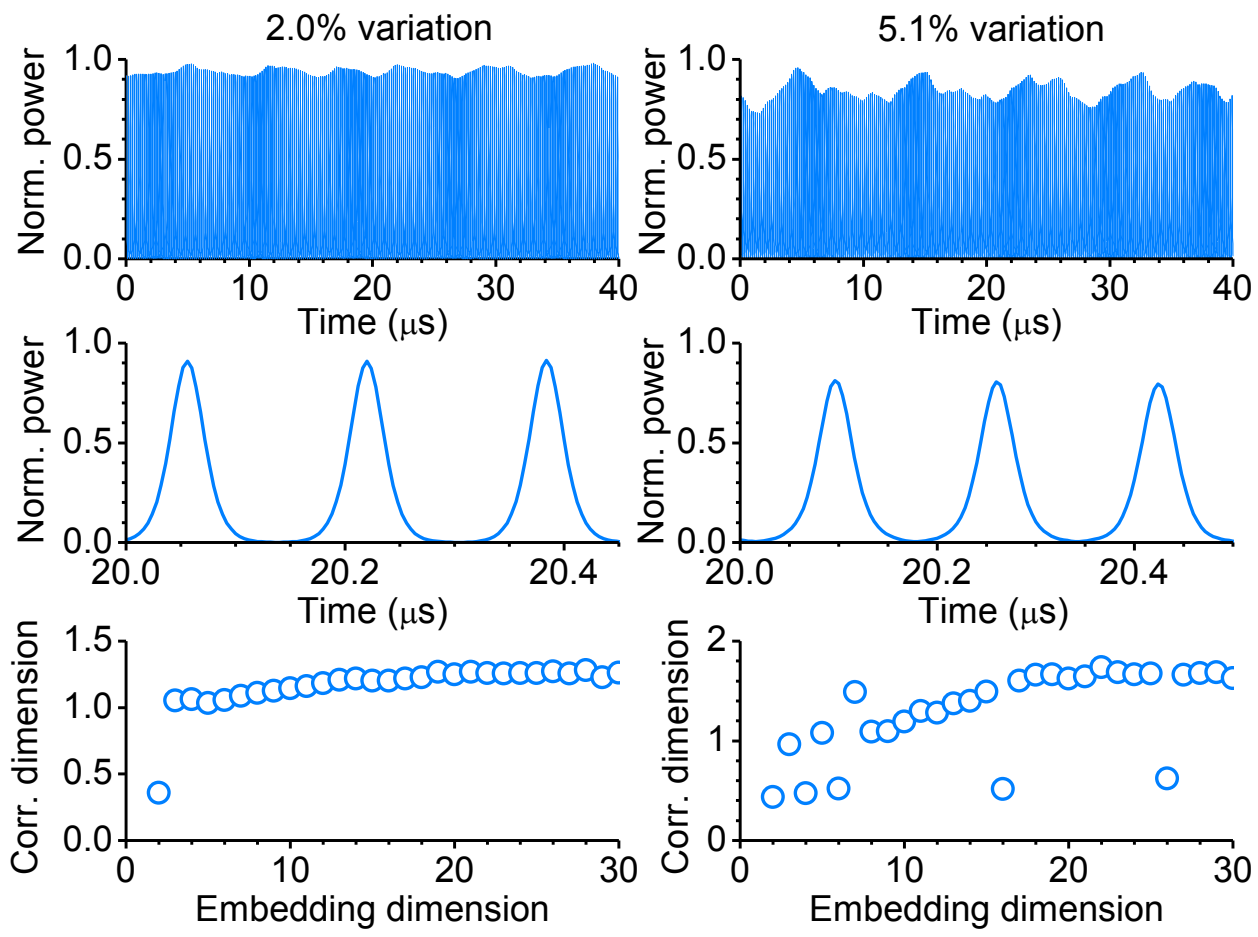

FIG. 4. Simulation results for chaotic spin-wave solitons.

In summary, this letter reports on the experimental observation of a spin-wave soliton that circulates in a magnetic film feedback ring with chaotically varying amplitude. The observed responses were reproduced by numerical simulations. There are two additional points of note. (1) There is a recent work on the propagation of a train of spin-wave solitons with amplitudes differing chaotically in a YIG strip [24]. The YIG strip in that work represents a 1D dissipative system, rather than the driven ring system considered in this letter. In this sense, the nonlinear objects in that work are significantly different from the chaotic soliton presented above. (2) The present work was realized in a regime where three-wave interactions were prohibited. Recent work has shown that soliton-like pulses with a chaotic phase modulation can develop in a regime where both three- and four-wave processes are allowed [25].

This work was supported by the U. S. National Science Foundation and the Russian Foundation for Basic Research.

*Corresponding author.

E-mail: mwu@lamar.colostate.edu